\documentclass[letterpaper, 10 pt, conference]{ieeeconf}

\IEEEoverridecommandlockouts                              

\usepackage{graphicx}      
\usepackage{cite}
\usepackage{grffile}
\usepackage{caption}
\usepackage{graphicx}
\usepackage{amssymb,amsmath}
\usepackage{epsfig}
\usepackage{epstopdf}
\usepackage{setspace}
\usepackage{amsthm}
\usepackage{enumerate}
\usepackage{tikz}
\usepackage{multirow}
\usepackage{algorithm}
\usepackage{algcompatible}
\usepackage{algpseudocode}
\usepackage{pifont}
\usetikzlibrary{shapes,arrows}
\title{\LARGE \bf
	Generalized Nash Equilibrium Problem by the Alternating Direction Method of Multipliers
}

\author{Farzad Salehisadaghiani, and Lacra Pavel
	\thanks{The authors are with the Department of Electrical and Computer Engineering, University of Toronto, Toronto, ON M5S 3G4, Canada (e-mails: {\tt\small farzad.salehisadaghiani@mail.utoronto.ca, pavel@ece.utoronto.ca}).}}
\newtheorem{assumption}{Assumption}

\newtheorem{theorem}{Theorem}

\newtheorem{definition}{Definition}

\begin{document}
		\allowdisplaybreaks

	\maketitle
	\thispagestyle{empty}
	\pagestyle{empty}

\begin{abstract}                
In this paper, the problem of finding a generalized Nash equilibrium (GNE) of a networked game is studied. Players are only able to choose their decisions from a feasible action set. The feasible set is considered to be a private linear equality constraint that is coupled through decisions of the other players. We consider that each player has his own private constraint and it has not to be shared with the other players. This general case also embodies the one with shared constraints between players and it can be also simply extended to the case with inequality constraints. Since the players don't have access to other players' actions, they need to exchange estimates of others' actions and a local copy of the Lagrangian multiplier with their neighbors over a connected communication graph. We develop a relatively fast algorithm by reformulating the conservative GNE problem within the framework of inexact-ADMM. The convergence of the algorithm is guaranteed under a few mild assumptions on cost functions. Finally, the algorithm is simulated for a wireless ad-hoc network.
\end{abstract}


\section{Introduction}
The problem of finding a generalized Nash equilibrium (GNE) has recently drawn many attentions due to its applicability in various networked games with coupling constraints such as power grids \cite{zhu2016distributed}, optical networks \cite{jayash10} and wireless communication networks \cite{han2012game}.

In such games, each player aims to minimize his cost function by taking a proper action in response to other players. Each player's feasible decision set is dependent on other players' actions. We are interested in seeking a GNE, which is a point that no player can unilaterally deviate his local action to minimize his cost function.

Due to similarities between this problem and distributed consensus optimization problems (DCOPs), we aim to employ an efficient, robust and fast optimization technique referred to as the {\emph{ alternating direction method of multipliers}} (ADMM) to find a GNE of a multi-player game. The key differences between DCOP and GNE problem is that in a DCOP each agent desires to minimize a global objective by controlling a full vector optimization variable. However, in a GNE problem, we have a set of local optimization problems assigned to each player who controls only his action (which is an element of a full vector). This leads to have a sub-optimization problem associated to each player such that each of them is dependent on the other players' actions.

\emph{\textbf{Related Works.}} Our work is related to the literature on (generalized) Nash games such as \cite{jayash8,facchinei2010generalized,fischer2014generalized,ssalehisadaghiani2016distributed} and DCOPs such as \cite{chang2015multi,johansson2008distributed}. (G)NE seeking in distributed networked games has recently attracted an increased interest due to many real-world applications. To name only a few, \cite{parise2015network,salehisadaghiani2016distributed}.

In \cite{parise2015network}, the problem of finding GNE is studied in a networked game. A gradient-based algorithm is designed over a complete communication graph. Convergence proof is analyzed in the presence of delay and dynamic change. Similar to \cite{zhu2016distributed}, the problem of seeking a GNE is discussed in \cite{yi2017distributed} for the games that players have access to all players' actions on which their cost depends on. Thus, they exchange only local multipliers not the estimates of others' actions.

A Nash game with a coupled constraint is considered in \cite{yin2011nash}. A variational inequality related approach is used to compute a GNE of the game. The authors in \cite{schiro2013solution} solve a quadratic GNE problem with linear shared coupled constraint using Lemkes method.

Recently, there has been a widespread research on GNE seeking in aggregative games \cite{zhu2016distributed,yin2011nash,schiro2013solution}. Quadratic aggregative games with affine and convex coupling constraints are studied in \cite{paccagnan2016distributed,grammatico2016aggregative}. A coordination scheme belonging to a class of asymmetric projection algorithm is presented in \cite{paccagnan2016distributed} and its convergence to a GNE is then discussed. A model-free dynamic control law, based on monotone operator, is proposed in \cite{grammatico2016aggregative} to ensure the global convergence to a GNE. 

ADMM algorithms, which are in the scope of this paper, have been developed in 70's to find an efficient way to obtain an optimal point of distributed optimization problems (DOPs) \cite{bertsekas1999parallel,boyd2011distributed}. After the re-introduction of this method in \cite{boyd2011distributed}, ADMM has become widely used to locate optimal points of DOPs as a robust and fast technique \cite{he20121,wei2012distributed}. An inexact-ADMM algorithm is proposed in \cite{chang2015multi} for DCOPs with an affine equality constraint. In \cite{salehisadaghiani2016distributedifac}, we develop a methodology to relate a distributed NE problem with no coupled constraint to a DCOP using augmentation technique. It is shown that an NE problem can be treated as a set of sub-optimization problems with an augmented equality constraint for the estimates of the players.

\emph{\textbf{Contributions.}} Motivated by ADMM methods designed for DCOPs, we develop a relatively fast algorithm for GNE problems within the framework of inexact-ADMM. Players are only aware of their own cost functions (which are not in the form of aggregative but general game), problem data (which is related to a private coupled equality constraint for each player) and action set of all players (they are not aware of the others' actions). We reformulate the conservative GNE problem into the corresponding Lagrange dual problem and then we augment it {\emph{using local estimates of players' actions as well as local copies of Lagrange multipliers}}. We derive a set of GNE conditions using the associated KKT conditions and finally we develop an inexact-ADMM algorithm based on the augmented Lagrange function which is related to each player. The convergence proof of the algorithm to a GNE of the game is then provided under a few mild assumptions on players' cost functions. 

The paper is organized as follows. The problem statement and assumptions are provided in Section~II. In Section~III, an inexact-ADMM-like algorithm is derived. Convergence of the proposed algorithm to a GNE of the game is discussed in Section~IV. Simulation results are illustrated in Section~V and concluding remarks are presented in Section~VI.
\vspace{-0.2cm} 
\section{Problem Statement}\label{problem_statement}

Consider $V=\{1,\ldots,N\}$ as a set of $N$ players that seek a GNE of a networked game with individual linear coupled constraints. The game is denoted by $\mathcal{G}$ and defined as follows:\vspace{-0.0cm}
\begin{itemize}
	\item $\Omega_i\subset\mathbb{R}$: Action set of player $i$, $\forall i\in V$,
	\item $\Omega=\prod_{i\in V}\Omega_i\subset\mathbb{R}^N$: Action set of all players,
	\item $J_i:\Omega\rightarrow \mathbb{R}$: Cost function of player $i$, $\forall i\in V$.
\end{itemize}

The game $\mathcal{G}(V,\Omega_i,J_i)$ is defined over the set of players, $V$, the action set of player $i\in V$, $\Omega_i$ and the cost function of player $i\in V$, $J_i$.

The players' actions are denoted as follows:\vspace{-0.0cm}
\begin{itemize}
	\item $x=(x_i,x_{-i})\in\Omega$: All players actions,
	\item $x_i\in\Omega_i$: Player $i$'s action, $\forall i\in V$,
	\item $x_{-i}\!\in\!\Omega_{-i}\!:=\!\prod_{j\in V\backslash\{i\}}\!\Omega_j$:All players' actions except $i$.	
\end{itemize}
The local data of player $i$ is given as $A^i:=[A_i^i,[A_j^i]_{j\neq i}]\in \mathbb{R}^{m\times N}$ with $A_i^i\in\mathbb{R}^m$, and $b^i\in\mathbb{R}^m$. Note that in this work, we are interested in a more general case in which each player has access only to his own private constraint which has not to be shared with the other players.
\begin{assumption}\label{constraint}
	\footnote[1]{Derivation of the algorithm is not dependent on Assumption~\ref{constraint}. Simulations for various cases for matrix $A^i$, even the ones that do not satisfy this assumption, verify that the algorithm works even without considering this assumption. However, this assumption is used at the convergence proof.}There exists a positive semi-definite matrix $B^i\in\mathbb{R}^{m\times m}$ for all $i\in V$ such that,
	\begin{equation}
	A_i^i\in \ker(B^i-I_m),\,A_j^i\in \ker(B^i)\quad j\neq i.	
	\end{equation}
\end{assumption}
Player $i$'s feasible strategy set is then denoted by $\mathcal{X}_i(x_{-i}):=\{x_i\in\Omega_i|A^ix=b^i\}$\footnote[2]{This case contains games with shared constraints for which $Ax=b$ for all $i\in V$ and $m\geq N$}.

The game is played such that for a given $x_{-i}\in \Omega_{-i}$, each player $i\in V$ aims to minimize his own cost function selfishly w.r.t. $x_i$ subject to a private equality constraint.
\begin{equation}
\label{mini_0}
\begin{cases}
\begin{aligned}
& \underset{x_i}{\text{minimize}}
& & J_i(x_i,x_{-i}) \\
& \hspace{0.5cm}\text{s.t.}
& & x_i\in\mathcal{X}_i(x_{-i}).
\end{aligned}
\end{cases}
\end{equation}
Each optimization problem is run by a particular player $i$ at the same time with other players. Note that \eqref{mini_0} can simply accommodate inequality constraints $A^ix\geq b^i$ by adding a slack variable $y\geq 0$ and rewrite the inequality constraint as $A^ix-y=b^i$. Note that the constraint $y\geq 0$ can be enforced by a convex indicator function.

An NE of a game is defined as follows:
\begin{definition}\label{Nash_def}
	Consider an $N$-player game $\mathcal{G}(V,\Omega_i,J_i)$, each player $i$ minimizing the cost function $J_i:\Omega\rightarrow\mathbb{R}$. A vector $x^*=(x_i^*,x_{-i}^*)\in\Omega$ is called a GNE of this game if
	\begin{equation}
	J_i(x_i^*,{x_{-i}^{*}})\leq J_i(x_{i},{x_{-i}^{*}})\quad\forall x_i\in \mathcal{X}_i(x_{-i}^*),\,\,\forall i\in V.
	\end{equation}
\end{definition}

An NE lies at the intersection of all solutions of the set \eqref{mini_0}.
The key challenges are two-fold. First, each optimization problem in \eqref{mini_0} is dependent on the solution of the other simultaneous problems; second, the solutions are coupled through a linear constraint. 

Since this game is distributed, no player is aware of the actions, cost functions and constraints of the other players. Thus, each player $i$ maintains an estimate of the other players' actions. We define a few notations, in the following,  for players' estimates.\vspace{-0cm}
\begin{itemize}
	\item $x^i=(x_i^i,x_{-i}^i)\in\Omega$: Player $i$'s estimate of all players actions,
	\item $x_i^i\in\Omega_i$: Player $i$'s estimate of his own action which is indeed his action, i.e., $x_i^i=x_i\,\forall i\in V$,
	\item $x_{-i}^i\in\Omega_{-i}:=\prod_{j\in V\backslash\{i\}}\Omega_j$: Player $i$'s estimate of all other players' actions except his action,
	\item $\underline{x}=[{x^1}^T,\ldots,{x^N}^T]^T\in\Omega^N$: Augmented vector of estimates of all players' actions
\end{itemize} 
Note that all players' actions can be interchangeably represented as $x=(x_i^i)_{i\in V}$.

We assume that the cost function $J_i$, the action set $\Omega$ and the problem data $A^i,\,b^i$ are the only information available to player $i$. Thus, the players exchange their estimates in order to update their actions. An undirected \emph{communication graph} $G_C(V,E)$ is defined with $E\subseteq V\times V$ denoting the set of communication links between the players. $(i,j)\in E$ if and only if players $i$ and $j$ exchange estimates. In the following, we have a few definitions for $G_C$:
\begin{itemize}\vspace{-0cm}
	\item $N_i:=\{j\in V|(i,j)\in E\}$: Set of neighbors of $i$ in $G_C$,
	\item $H:=[h_{ij}]_{i,j\in V}$: Adjacency matrix associated with $G_C$ where $h_{ij}=1$ if $(i,j)\in E$ and $h_{ij}=0$ otherwise,
	\item $D:=\text{diag}\{|N_1|,\ldots,|N_N|\}$: Degree matrix associated with $G_C$. 
\end{itemize} 
The following assumption is used for $G_C$.  
\begin{assumption}\label{connectivity}
	$G_C$ is a connected graph.
\end{assumption}

We aim to link game \eqref{mini_0} to a set of optimization problems whose solutions can be based on the ADMM. 
Game \eqref{mini_0} is equivalently represented as the following problem for $i\in V$.
\begin{equation}\label{mini_11}
\begin{cases}
\begin{aligned}
& \underset{x_i\in\Omega_i}{\text{min}}
& & J_i(x_i,x_{-i}) \\
& \hspace{0.2cm}\text{s.t.}
& & A^ix=b^i.
\end{aligned}
\end{cases}
\end{equation}
Let $\lambda\in\mathbb{R}^m$ be the joint Lagrange dual variable associated with the linear constraint $A^ix=b^i\,\forall i\in V$. The Lagrange dual problem of \eqref{mini_11} can be written as follows for $i\in V$.
\begin{equation}\label{mini_22}
\begin{cases}
\begin{aligned}
& \underset{\lambda\in\mathbb{R}^m}{\text{max}}\underset{x_i\in\Omega_i}{\text{min}}
& & J_i(x_i,x_{-i})+\lambda^T(A^ix-b^i).
\end{aligned}
\end{cases}
\end{equation}
Note that we are interested in computing a variational GNE of the game since we assumed a common Lagrange multiplier $\lambda$ for each player in \eqref{mini_22}. Using the estimates of the actions $x^i$ and the local copies of the Lagrange dual variables $\lambda^i$ for $i\in V$, we reformulate \eqref{mini_22} so that the objective function is separable (the estimates are also interpreted as the local copies of $x$). Moreover, the slack variables $t^{ls}|_{\{l\in V,\,s\in N_l\}}\in\Omega$ and $p^{ls}|_{\{l\in V,\,s\in N_l\}}\in\mathbb{R}^m$ are employed to enforce that the local copies are equivalent.
\begin{equation}\label{mini_33}
\hspace{-0cm}\begin{cases}
\begin{aligned}
& \underset{\lambda^i\in\mathbb{R}^m,p^{ls}}{\text{max}}\,\underset{x_i^i\in\Omega_i,t^{ls}}{\text{min}}
& & J_i\!(\!x_i^i,x_{-i}^i)\!+\!\mathcal{I}_{\Omega_i}(x_i^i)\!+\!{\lambda^i}^T(A^ix^i\!-\!b^i)\\
& \hspace{1cm}\text{s.t.}
& & x^l=t^{ls}\quad\forall l\in V,\,\forall s\in N_l,\\
& & & x^s=t^{ls}\quad\forall l\in V,\,\forall s\in N_l,\\
& & & \lambda^l=p^{ls}\quad\forall l\in V,\,\forall s\in N_l,\\
& & & \lambda^s=p^{ls}\quad\forall l\in V,\,\forall s\in N_l,
\end{aligned}
\end{cases}
\end{equation}
where $\mathcal{I}_{\Omega_i}(x_i^i):=\begin{cases}0&\text{if }x_i^i\in\Omega_i\\\infty&\text{otherwise}\end{cases}$ is an indicator function of the feasibility constraint $x_i^i\in\Omega_i$. 
Note that the solution of \eqref{mini_33} is equivalent to that of \eqref{mini_22}, i.e., $x_i^i=x_i,\,\lambda^i=\lambda$ for all $i\in V$. The ADMM is then employed to solve \eqref{mini_33} in a distributed manner. A characterization of the NE for game \eqref{mini_0} could be obtained by finding the KKT conditions on \eqref{mini_33}. Let $\{u^{ls},v^{ls},w^{ls},z^{ls}\}_{l\in V,s\in N_l}$ with $u^{ls},v^{ls}\in \mathbb{R}^N$ and $w^{ls},z^{ls}\in \mathbb{R}^m$ be the Lagrange multipliers associated with the four constraints in \eqref{mini_33}, respectively. The corresponding Lagrange function for player $i$, $\forall i\in V$ is as follows:
\begin{eqnarray}
&&\hspace{-0.7cm}L_i\Big(x_i^i, \{t^{ls}, u^{ls}, v^{ls}\},\lambda^i, \{p^{ls}, w^{ls}, z^{ls}\}\Big)\nonumber\\
&&\hspace{-0.7cm}:= J_i(x_i^{i},x_{-i}^{i})+\mathcal{I}_{\Omega_i}(x_i^{i})+{\lambda^{i}}^T(A^ix^{i}-b^i)\nonumber\\
&&\hspace{-0.7cm}+\sum_{l\in V}\sum_{s\in N_l}{u^{ls}}^T(x^l-t^{ls})+{v^{ls}}^T(x^s-t^{ls}),\nonumber\\
&&\hspace{-0.7cm}-\sum_{l\in V}\sum_{s\in N_l}{w^{ls}}^T(\lambda^l-p^{ls})+{z^{ls}}^T(\lambda^s-p^{ls}).
\end{eqnarray}
Let $({{x}^{i}}^*, {{{\lambda}^{i}}^*})_{i\in V}$ and $\{{{u}^{ls}}^*,{{v}^{ls}}^*,{{w}^{ls}}^*,{{z}^{ls}}^*\}_{l\in V,\,s\in N_l}$ be a pair of primal and dual optimal solution to \eqref{mini_33}. Let denote $\sum_{j\in N_i}u^{ij}+v^{ji}:=U^i$ and $\sum_{j\in N_i}w^{ij}+z^{ji}:=W^i$ for notational simplicity. The KKT conditions are summarized as follows for $i\in V,\,j\in N_i$:\vspace{-0cm}
\begin{eqnarray}\label{kkt_dual}
\hspace{-0.8cm}\begin{cases}\nabla_iJ_i({{x}^{i}}^*)+\partial\mathcal{I}_{\Omega_i}({{x}_i^{i}}^*)+{\lambda^i}^{*T}A_i^i+{U_i^i}^*=0,\\
A^i{{x}^{i}}^*-b^i-{W^i}^*=\textbf{0}_m,\\
{{x}^{i}}^*={{x}^{j}}^*,\,{{\lambda}^{i}}^*={{\lambda}^{j}}^*,\\
{{u}^{ij}}^*+{{v}^{ij}}^*=\textbf{0}_N,\,{{w}^{ij}}^*+{{z}^{ij}}^*=\textbf{0}_m,
\end{cases}
\end{eqnarray}
where $\nabla_iJ_i(\cdot)$ is gradient of $J_i$ w.r.t. $x_i$ and $\partial_i\mathcal{I}_{\Omega_i}(\cdot)$ is a subgradient of $\mathcal{I}_{\Omega_i}$ at $x_i$. By \eqref{kkt_dual} and Assumption~\ref{connectivity}, ${{x}^{1}}^*\!=\!\ldots\!=\!{{x}^{N}}^*\!:=\!x^*$ and ${{\lambda}^{1}}^*\!=\!\ldots\!=\!{{\lambda}^{N}}^*\!:=\!\lambda^*$. Then, the GNE of \eqref{mini_0}, $x^*$, satisfies the following equations $\forall i\in V,\,j\in N_i$:
\begin{eqnarray}\label{Nash_equations}
\begin{cases}
\nabla_iJ_i({x}^*)+\partial\mathcal{I}_{\Omega_i}({x_i}^*)+{\lambda^{*}}^TA_i^i+{U_i^i}^*=0,\\
A^ix^{*}-b^i-{W^i}^*=\textbf{0}_m,\\
{u^{ij}}^*+{v^{ij}}^*=\textbf{0}_N,\,{w^{ij}}^*+{z^{ij}}^*=\textbf{0}_m.
\end{cases}
\end{eqnarray}	 
In the following, we state a few assumptions for the existence of a GNE.
\begin{assumption}
	\label{assump}
	For every $i\in V$, 
	\begin{itemize}
	\item $\Omega_i\subset\mathbb{R}$ is compact and convex
	\item $\mathcal{X}_i(x_{-i})$ is non-empty for every $x_{-i}$	
	\item $J_i(x_i,x_{-i})$ is $C^1$ in $x_i$, jointly continuous in $x$ and convex in $x_i$, for every $x_{-i}$.
	\end{itemize}
\end{assumption}
The convexity of $\Omega_i$ implies that the indicator function  $\mathcal{I}_{\Omega_i}$ is convex. This yields that there exists at least one bounded subgradient $\partial\mathcal{I}_{\Omega_i}$. 
\begin{assumption}\label{Lip_assump}
	Let $F:\Omega^N\rightarrow\mathbb{R}^N$, $F(\underline{x}):=[\nabla_iJ_i(x^i)]_{i\in V}$ be the pseudo-gradient vector (game map) where $\underline{x}:=[{x^1}^T,\ldots,{x^N}^T]^T\in\Omega^N$. $F$ is cocoercive $\forall\underline{x}\in\Omega^N$ and $\underline{y}\in\Omega^N$, i.e.,\vspace{-0.4cm}
	\begin{eqnarray}
	&&\hspace{-1cm}(F(\underline{x})-F(\underline{y}))^T(x-y)\geq \sigma_F\|F(\underline{x})-F(\underline{y})\|^2,
	\end{eqnarray}
	where $\sigma_F>0$.
\end{assumption}
\vspace{-0.4cm}
\section{Distributed Inexact-ADMM Algorithm}
Our objective is to find an ADMM-like 
algorithm for computing a GNE of $\mathcal{G}(V,\Omega_i,J_i)$ using only imperfect information over the communication graph $G_C(V,E)$.

The algorithm is elaborated in the following steps:\\
1- \textbf{\emph{Initialization Step:}}
Each player $i\in V$ maintains an initial estimate for all players' actions, $x^i(0)\in\Omega$ and an initial Lagrange multiplier, $\lambda^i(0)\in\mathbb{R}^m$. The initial values of $u^{ij}(0),v^{ij}(0),w^{ij}(0)$ and $z^{ij}(0)$ are all set to be zero for all $i\in V$, $j\in N_i$.\\
2- \textbf{\emph{Communication Step:}}
At iteration $T(k)$, each player $i\in V$ exchanges his previous estimate of the other players' actions $x^i(k-1)$ and his dual Lagrange multiplier $\lambda^i(k-1)$ with his neighbors $j\in N_i$.\\
3- \textbf{\emph{Action Update Step:}} At this moment all players update their actions, estimates and Lagrange multiplier via the ADMM. For each player $i\in V$, let the augmented Lagrange function associated with \eqref{mini_33} be as follows:
\begin{eqnarray}\label{Augmented_Lag}
&&\hspace{-1cm}L_i^a\Big(x_i^i, \{t^{ls}, u^{ls}, v^{ls}\},\lambda^i, \{p^{ls}, w^{ls}, z^{ls}\}\Big)\nonumber\\
&&\hspace{-1cm}:= J_i(x_i^{i},x_{-i}^{i})+\mathcal{I}_{\Omega_i}(x_i^{i})+{\lambda^{i}}^T(A^ix^{i}-b^i)\nonumber\\
&&\hspace{-1cm}+\sum_{l\in V}\sum_{s\in N_l}{u^{ls}}^T(x^l-t^{ls})+{v^{ls}}^T(x^s-t^{ls})\nonumber\\
&&\hspace{-1cm}-\sum_{l\in V}\sum_{s\in N_l}{w^{ls}}^T(\lambda^l-p^{ls})+{z^{ls}}^T(\lambda^s-p^{ls})\nonumber\\
&&\hspace{-1cm}+\frac{c}{2}\sum_{l\in V}\sum_{s\in N_l}(\|x^l-t^{ls}\|^2+\|x^s-t^{ls}\|^2)\nonumber\\
&&\hspace{-1cm}-\frac{c}{2}\sum_{l\in V}\sum_{s\in N_l}(\|\lambda^l-p^{ls}\|^2+\|\lambda^s-p^{ls}\|^2),
\end{eqnarray}
where $c>0$ is a scalar coefficient. Consider the ADMM algorithm associated with \eqref{mini_33} based on \eqref{Augmented_Lag}:

The dual Lagrange multipliers $u^{ij},v^{ij},w^{ij},z^{ij}$ update rules $\forall i\in V,j\in N_i$ are as follows:\vspace{-0cm}
\begin{eqnarray}
&&\hspace{-1cm}u^{ij}(k)=u^{ij}(k-1)+\frac{c}{2}\big(x^i(k-1)-x^j(k-1)\big),\label{u_update_ADMM}\\
&&\hspace{-1cm}v^{ij}(k)=v^{ij}(k-1)+\frac{c}{2}\big(x^j(k-1)-x^i(k-1)\big),\label{v_update_ADMM}\\
&&\hspace{-1cm}w^{ij}(k)=w^{ij}(k-1)+\frac{c}{2}\big(\lambda^i(k-1)-\lambda^j(k-1)\big),\label{w_update_ADMM}\\
&&\hspace{-1cm}z^{ij}(k)=z^{ij}(k-1)+\frac{c}{2}\big(\lambda^j(k-1)-\lambda^i(k-1)\big).\label{z_update_ADMM}	
\end{eqnarray}

The update rule for the slack variable $t^{ij}\in\mathbb{R}^N$, $\forall i\in V, j\in N_i$, which is based on \eqref{Augmented_Lag}, is as follows:\vspace{-0cm}
\begin{eqnarray}\label{tijavalesh}
&&\hspace{-0.7cm}t^{ij}(k)=\text{arg }\min_{t^{ij}} L_i^{a}\Big(x^i(k-1), \{t^{ls},u^{ls}(k),v^{ls}(k)\},\nonumber\\
&&\hspace{-0.7cm}\lambda^i(k-1),\{p^{ls}(k-1),w^{ls}(k),z^{ls}(k)\}\Big)\nonumber\\
&&\hspace{-0.7cm}=\text{arg }\min_{t^{ij}}\Big\{-(u^{ij}(k)+v^{ij}(k))^Tt^{ij}\nonumber\\
&&\hspace{-0.7cm}+\frac{c}{2}(\|x^i(k-1)-t^{ij}\|^2+\|x^j(k-1)-t^{ij}\|^2)\Big\}\nonumber\\
&&\hspace{-0.7cm}=\frac{1}{2c}(u^{ij}(k)+v^{ij}(k))+\frac{1}{2}(x^i(k-1)+x^j(k-1)).\label{x_i^i_ADMM}
\end{eqnarray}
The initial conditions $u^{ij}(0)=v^{ij}(0)=\textbf{0}_N$ $\forall i\in V,\,j\in N_i$ along with \eqref{u_update_ADMM} and \eqref{v_update_ADMM} suggest that $u^{ij}(k)+v^{ij}(k)=\textbf{0}_N$ $\forall i\in V,\,j\in N_i,\,k>0$. Then, \eqref{tijavalesh} yields,
\begin{equation}
t^{ij}(k)=\frac{x^i(k-1)+x^j(k-1)}{2}.\label{t_ij_simp}
\end{equation}

Similarly the update rule for $p^{ij}\in\mathbb{R}^m$ $\forall i\in V, j\in N_i$ is represented as the following:\vspace{-0cm}
\begin{equation}
p^{ij}(k)=\frac{\lambda^i(k-1)+\lambda^j(k-1)}{2}.\label{p_ij_simp}
\end{equation}
The local estimate update for all $i\in V$ is a min-max optimization problem based on \eqref{Augmented_Lag}.
\begin{eqnarray}\label{x_i^i start}
&&\hspace{-0.8cm}x_i^i(k)\!=\!\text{arg }\min_{x_i^i\in\mathbb{R}}\max_{\lambda^i\in\mathbb{R}^m}\!L_i^{a}\Big(x_i^i, \{t^{ls}(k), u^{ls}(k), v^{ls}(k)\},\lambda^i,\nonumber\\
&&\hspace{-0.8cm}\{p^{ls}(k), w^{ls}(k), z^{ls}(k)\}\Big)\nonumber\\
&&\hspace{-0.8cm}=\text{arg }\min_{x_i^i}\max_{\lambda^i}\Big\{J_i(x_i^i,x_{-i}^i(k-1))+\mathcal{I}_{\Omega_i}(x_i^i)\nonumber\\
&&\hspace{-0.8cm}-{\lambda^i}^Tb^i+{\lambda^i}^TA_i^ix_i^i+{\lambda^i}^TA_{-i}^ix_{-i}^i(k-1)\nonumber\\
&&\hspace{-0.8cm}+{U_i^i}(k)^Tx_i^i-{W^i}(k)^T\lambda^i\nonumber\\
&&\hspace{-0.8cm}+c\sum_{j\in N_i}\Big\|x_i^i-\frac{x_i^i(k-1)+x_i^j(k-1)}{2}\Big\|^2\nonumber\\
&&\hspace{-0.8cm}-c\sum_{j\in N_i}\Big\|\lambda^i-\frac{\lambda^i(k-1)+\lambda^j(k-1)}{2}\Big\|^2\Big\}\quad\forall i\in V.\label{x_i^i_ADMM_bef}
\end{eqnarray}
In the derivation of \eqref{x_i^i start}, we used \eqref{t_ij_simp} and \eqref{p_ij_simp}. 
The next min-max problem is equivalent in the sense that its solutions for $x_i^i$ and $\lambda^i$ are equivalent to the solutions of \eqref{x_i^i start}. 
\begin{eqnarray}\label{x_i^ivasat}
&&\hspace{-0.8cm}x_i^i(k)=\text{arg }\min_{x_i^i}\max_{\lambda^i}\Big\{J_i(x_i^i,x_{-i}^i(k-1))+\mathcal{I}_{\Omega_i}(x_i^i)\nonumber\\
&&\hspace{-0.8cm}+c\sum_{j\in N_i}\Big\|x_i^i-\frac{x_i^i(k-1)+x_i^j(k-1)}{2}\Big\|^2+{U_i^i}(k)^Tx_i^i\\
&&\hspace{-0.8cm}+{\lambda^i}^T\Big(A_i^ix_i^i+A_{-i}^ix_{-i}^i(k-1)-b^i-{W^i}(k)\nonumber\\
&&\hspace{-0.8cm}+c\sum_{j\in N_i}(\lambda^i(k-1)+\lambda^j(k-1))\Big)-c|N_i|\|\lambda^i\|^2\Big\}\quad\forall i\in V.\label{x_i^i_ADMM_bef}\nonumber
\end{eqnarray}
For the last two lines of \eqref{x_i^ivasat}, we complete the squared term as follows:
\begin{eqnarray}\label{x_i^ifaghatlambda^i}
&&\hspace{-0.2cm}x_i^i(k)=\text{arg }\min_{x_i^i}\max_{\lambda^i}\Big\{J_i(x_i^i,x_{-i}^i(k-1))+\mathcal{I}_{\Omega_i}(x_i^i)\nonumber\\
&&\hspace{-0.2cm}+c\sum_{j\in N_i}\Big\|x_i^i-\frac{x_i^i(k-1)+x_i^j(k-1)}{2}\Big\|^2+{U_i^i}(k)^Tx_i^i\nonumber\\
&&\hspace{-0.2cm}-c|N_i|\Big\|\lambda^i-\frac{1}{2c|N_i|}\Big[c\sum_{j\in N_i}(\lambda^i(k-1)+\lambda^j(k-1))\nonumber\\
&&\hspace{-0.2cm}-{W^i}(k)+A_i^ix_i^i+A_{-i}^ix_{-i}^i(k-1)-b^i\Big]\Big\|^2\nonumber\\
&&\hspace{-0.2cm}+\frac{1}{4c|N_i|}\Big\|c\sum_{j\in N_i}(\lambda^i(k-1)+\lambda^j(k-1))\nonumber\\
&&\hspace{-0.2cm}-{W^i}(k)+A_i^ix_i^i+A_{-i}^ix_{-i}^i(k-1)-b^i\Big\|^2\Big\}.
\end{eqnarray}
Only the third line of \eqref{x_i^ifaghatlambda^i} is dependent on $\lambda^i$ which is the maximization variable. Thus, to maximize \eqref{x_i^ifaghatlambda^i} w.r.t. $\lambda^i$, we let $\lambda^i(k)$ be as follows:
\begin{eqnarray}\label{lambda^i}
&&\hspace{-1cm}\lambda^i(k)=\frac{1}{2c|N_i|}\Big(A_i^ix_i^{i}(k)+A_{-i}^ix_{-i}^{i}(k-1)-b^i\nonumber\\
&&\hspace{-1cm}-{W^i}(k)+2c\sum_{j\in N_i}\frac{\lambda^i(k-1)+\lambda^j(k-1)}{2}\Big).
\end{eqnarray} 
Substituting back \eqref{lambda^i} into \eqref{x_i^ifaghatlambda^i}, we obtain,
\begin{eqnarray}\label{x_i^ighableproximation}
&&\hspace{-0.2cm}x_i^i(k)=\text{arg }\min_{x_i^i}\Big\{J_i(x_i^i,x_{-i}^i(k-1))+\mathcal{I}_{\Omega_i}(x_i^i)\nonumber\\
&&\hspace{-0.2cm}+c\sum_{j\in N_i}\Big\|x_i^i-\frac{x_i^i(k-1)+x_i^j(k-1)}{2}\Big\|^2+{U_i^i}(k)^Tx_i^i\nonumber\\
&&\hspace{-0.2cm}+\frac{1}{4c|N_i|}\Big\|c\sum_{j\in N_i}(\lambda^i(k-1)+\lambda^j(k-1))\nonumber\\
&&\hspace{-0.2cm}-{W^i}(k)+A_i^ix_i^i+A_{-i}^ix_{-i}^i(k-1)-b^i\Big\|^2\Big\}.
\end{eqnarray}
We simplify \eqref{x_i^ighableproximation} by using a proximal first-order approximation for $J_i(x_i^i,x_{-i}^i(k-1))$ and the last term in \eqref{x_i^ighableproximation} around $x^i(k-1)$; thus  using inexact ADMM it follows:
\begin{eqnarray}\label{x_i^ighableakhar}
&&\hspace{-0.7cm}x_i^i(k)=\text{arg }\min_{x_i^i}\Big\{\Big[\nabla_iJ_i(x^i(k-1))+\frac{{A_i^i}^T}{2c|N_i|}\Big(A^ix^i(k-1)\nonumber\\
&&\hspace{-0.7cm}-b^i-{W^i}(k)+c\sum_{j\in N_i}(\lambda^i(k-1)+\lambda^j(k-1))\Big)\Big]^T.\nonumber\\
&&\hspace{-0.7cm}.(x_i^i-x_{i}^i(k-1))+\frac{\beta_i}{2}\|x_i^i-x_{i}^i(k-1)\|^2
+\mathcal{I}_{\Omega_i}(x_i^i)\nonumber\\
&&\hspace{-0.7cm}+c\sum_{j\in N_i}\Big\|x_i^i-\frac{x_i^i(k-1)+x_i^j(k-1)}{2}\Big\|^2+{U_i^i}(k)^Tx_i^i\Big\},
\end{eqnarray}
where $\beta_i>0$ is a penalty factor for the proximal first-order approximation for $i\in V$. 
We obtain an equivalent problem to \eqref{x_i^ighableakhar} in the following such that its gradient w.r.t. $x_i^i$ is equivalent to that of \eqref{x_i^ighableakhar}. 
\begin{eqnarray}\label{x_i^ioptimal}
&&\hspace{-0.7cm}x_i^i(k)=\text{arg }\min_{x_i^i}\Big\{\mathcal{I}_{\Omega_i}(x_i^i)+\frac{\alpha_i}{2}\Big\|x_i^i-\alpha_i^{-1}\Big(\beta_ix_i^i(k-1)\nonumber\\
&&\hspace{-0.7cm}-\nabla_iJ_i(x^i(k-1))-{U_i^i}(k)+c\sum_{j\in N_i}(x_i^i(k-1)+x_i^j(k-1))\nonumber\\
&&\hspace{-0.7cm}-\frac{{A_i^i}^T}{2c|N_i|}\Big(A^ix^{i}(k-1)-b^i-{W^i}(k)\nonumber\\
&&\hspace{-0.7cm}+c\sum_{j\in N_i}\lambda^i(k-1)+\lambda^j(k-1)\Big)\Big)\Big\|^2\Big\},
\end{eqnarray}
\vspace{-0cm}
where $\alpha_i:=\beta_i+2c|N_i|$. Let $\text{prox}_{g}^{a}[s]:=\text{arg }\min_{x}\{g(x)+\frac{a}{2}\|x-s\|^2\}$ be the proximal operator for the non-smooth function $g$. Note that $\text{prox}_{\mathcal{I}_{\Omega_i}}^{\alpha_i}[s]=T_{\Omega_i}[s]$ where $T_{\Omega_i}:\mathbb{R}\rightarrow\Omega_i$ is an Euclidean projection. Then for each player $i$, $\forall i\in V$ we obtain,
\vspace{-0cm}
\begin{eqnarray}
&&\hspace{-0.7cm}x_i^i(k)=T_{\Omega_i}\Big[\alpha_i^{-1}\Big(\gamma_ix_i^i(k-1)-\nabla_iJ_i(x^i(k-1))-{U_i^i}(k)\nonumber\\
&&\hspace{-0.7cm}+c\sum_{j\in N_i}x_i^j(k-1)-\frac{{A_i^i}^T}{2c|N_i|}\Big(A_{-i}^ix_{-i}^{i}(k-1)-b^i-{W^i}(k)\nonumber\\
&&\hspace{-0.7cm}+c\sum_{j\in N_i}\lambda^i(k-1)+\lambda^j(k-1)\Big)\Big)\Big],
\end{eqnarray}

where $\gamma_i:=\beta_i+c|N_i|-\frac{1}{2c|N_i|}{A_i^i}^TA_i^i$.

Eventually, after updating his action and dual Lagrange multiplier, each player $i$ takes average of the received information with his estimate and updates his estimate as follows:\vspace{-0cm}
\begin{equation}
x_{-i}^i(\!k\!)\!=\!\frac{1}{2}\!\Big(\!x_{-i}^i(\!k\!-\!1\!)\!+\!\underbrace{\frac{1}{|N_i|}\!\sum_{j\in N_i}\!x_{-i}^j\!(\!k\!-\!1\!)}_{\text{RECEIVED INFORMATION}}\Big)\!-\underbrace{\frac{1}{2c|N_i|}\!{U_{-i}^i}\!(k)}_{\text{PENALTY TERM}},\label{x_-i_Update}
\end{equation}
where $c>0$ is a scalar coefficient which is also used in \eqref{Augmented_Lag}. Note that in \eqref{x_-i_Update}, the penalty term is associated with the difference between the estimates of the neighboring players (Equations~\eqref{u_update_ADMM}, \eqref{v_update_ADMM}).

Then the ADMM algorithm is represented as in Algorithm~1.
\begin{algorithm}
		\caption{ADMM Algorithm}
	\label{ADMMalgorithm}
	\begin{algorithmic}[1]
		\State \textbf{initialization} $x^i(0)\in\Omega$, $\lambda^i(0)\in\mathbb{R}^m$, $U^{i}(0)=\textbf{0}_N$ and $W^{i}(0)=\textbf{0}_m$ $\forall i\in V$
		\For{$k=1,2,\ldots$ }
		\For{each player $i\in V$ }
		\State players $i$, $j$ $\forall j\in N_i$ exchange $x^i(k-1)$, 
		\Statex \hspace{0.95cm}
		$x^j(k-1)$, $\lambda^i(k-1)$ and $\lambda^j(k-1)$.
		\State $U^i(k)\!=\!U^i(k-1)\!+\!c\sum_{j\in N_i}(x^i(k-1)-x^j(k-1))$
		\State $W^i(k)\!=\!W^i(k-1)\!+\!c\sum_{j\in N_i}(\lambda^i(k-1)\!-\!\lambda^j(k\!-\!1))$
		\State $x_i^i(k)=T_{\Omega_i}\Big[\alpha_i^{-1}\Big(\gamma_ix_i^i(k-1)-\nabla_iJ_i(x^i(k-1))$
		\Statex\hspace{2.3cm}$-{U_i^i}(k)+c\sum_{j\in N_i}x_i^j(k-1)$
		\Statex\hspace{2.3cm}$-\frac{{A_i^i}^T}{2c|N_i|}\Big(A_{-i}^ix_{-i}^{i}(k-1)-b^i-{W^i}(k)$
		\Statex\hspace{2.3cm}
		$+c\sum_{j\in N_i}\lambda^i(k-1)+\lambda^j(k-1)\Big)\Big)\Big]$
		\State $\lambda^i(k)=\frac{1}{2c|N_i|}\Big(A_i^ix_i^{i}(k)+A_{-i}^ix_{-i}^{i}(k-1)-b^i$
		\Statex\hspace{2.3cm}$-{W^i}(k)+2c\sum_{j\in N_i}\frac{\lambda^i(k-1)+\lambda^j(k-1)}{2}\Big)$
		\State $x_{-i}^i(k)=\frac{\sum_{j\in N_i}x_{-i}^j(k-1))}{|N_i|}-\frac{U_{-i}^i(k-1)}{2c|N_i|}$
		\EndFor
		\EndFor
	\vspace{-0cm}
	\end{algorithmic}
	\vspace{-0cm}
\end{algorithm}
\vspace{-0.5cm}
\section{Convergence Proof}\label{convergence_proof}
\begin{theorem}\label{theorem_convergence_rate}
	Let $\beta_i>0$ be player $i$'s penalty factor of the approximation in the inexact ADMM Algorithm~\ref{ADMMalgorithm} which satisfies\vspace{-0.2cm}
	\begin{equation}\label{condition}
	\sigma_F>\frac{1}{2\beta_{i}-\frac{\|A_i^i\|^2}{c|N_i|}},
	\end{equation}
	where $\sigma_F$ is a positive constant for the cocoercive property of $F$, $A_i^i$ is player $i$'s problem data and $N_i$ is the number of neighbors of player $i$. Under Assumptions~\ref{constraint}-\ref{Lip_assump}, the sequence $\{x^i(k)\}$ $\forall i\in V$, generated by Algorithm~\ref{ADMMalgorithm} converges to $x^*$ NE of game \eqref{mini_0}. 
\end{theorem}
\par{\emph{Proof}}.
From step~7 in Algorithm~1, we obtain the following optimality condition:
\begin{eqnarray}\label{optmal}
&&\hspace{-0.2cm}\nabla_iJ_i(x^i(k-1))+\beta_i(x_i^i(k)-x_{i}^i(k-1))
+\partial_i\mathcal{I}_{\Omega_i}(x_i^i(k))\nonumber\\
&&\hspace{-0.2cm}+\frac{{A_i^i}^T}{2c|N_i|}\Big(A^ix^{i}(k-1)-b^i-W^i(k)\\
&&\hspace{-0.2cm}+c\sum_{j\in N_i}(\lambda^i(k-1)+\lambda^j(k-1))\Big)+U_i^i(k)\nonumber\\
&&\hspace{-0.2cm}+2c\sum_{j\in N_i}\Big(x_i^i(k)-\frac{x_i^i(k-1)+x_i^j(k-1)}{2}\Big)=0\quad\forall i\in V,\nonumber
\end{eqnarray}
Adding and subtracting $A_i^ix_i^i(k)$ from the second line of \eqref{optmal} and using \eqref{lambda^i}, it leads to
\begin{eqnarray}\label{darakhar1}
&&\hspace{-0.8cm}\nabla_iJ_i(x^i(k-1))
+\partial_i\mathcal{I}_{\Omega_i}(x_i^i(k))+{A_i^i}^T\lambda^i(k)\nonumber\\
&&\hspace{-0.8cm}+\delta_i(x_i^i(k)-x_i^i(k-1))+U_i^i(k)\\
&&\hspace{-0.8cm}+2c\sum_{j\in N_i}\Big(x_i^i(k)-\frac{x_i^i(k-1)+x_i^j(k-1)}{2}\Big)=0\quad\forall i\in V,\nonumber\vspace{-0cm}
\end{eqnarray}
where $\delta_i:=\beta_i-\frac{1}{2c|N_i|}{A_i^i}^TA_i^i$. Adding the first Nash condition \eqref{Nash_equations} and multiplying by $(x_i^i(k)-x_i^*)$, we obtain,
\begin{eqnarray}\label{optimalx_inahayi}
&&\hspace{-0.7cm}\Big(\nabla_iJ_i(x^i(k-1))-\nabla_iJ_i(x^*)\Big)^T(x_i^i(k-1)-x_i^*)\nonumber\\
&&\hspace{-0.7cm}+\Big(\nabla_iJ_i(x^i(k-1))-\nabla_iJ_i(x^*)\Big)^T(x_i^i(k)-x_i^i(k-1))\nonumber\\
&&\hspace{-0.7cm}+\Big(\partial_i\mathcal{I}_{\Omega_i}(x_i^i(k))-\partial_i\mathcal{I}_{\Omega_i}(x_i^*)\Big)^T(x_i^i(k)-x_i^*)\nonumber\\
&&\hspace{-0.7cm}+\Big(\lambda^i(k)-\lambda^*\Big)^TA_i^i(x_i^i(k)-x_i^*)\\
&&\hspace{-0.7cm}+\delta_i(x_i^i(k)-x_i^i(k-1))^T(x_i^i(k)-x_i^*)\nonumber\\
&&\hspace{-0.7cm}+\Big(U_i^i(k)-{U_i^i}^*\Big)^T(x_i^i(k)-x_i^*)\nonumber\\
&&\hspace{-0.7cm}+2c\!\sum_{j\in N_i}\!\Big(\!x_i^i(k)\!-\!\frac{x_i^i(k-1)\!+\!x_i^j(k-1)}{2}\!\Big)^T\!(\!x_i^i(k)\!-\!x_i^*)\!=\!0\nonumber
\end{eqnarray}

On the other hand, from \eqref{lambda^i} we obtain,
\begin{eqnarray}\label{darakhar2}
&&\hspace{-0.5cm}\textbf{0}_m=-A_i^ix_i^{i}(k)-A_{-i}^ix_{-i}^{i}(k-1)+b^i\nonumber\\
&&\hspace{-0.5cm}+W^i(k)+2c\sum_{j\in N_i}\lambda^i(k)-\frac{\lambda^i(k)+\lambda^j(k)}{2}\nonumber\\
&&\hspace{-0.5cm}+c\sum_{j\in N_i}(\lambda^i(k)+\lambda^j(k)-\lambda^i(k-1)-\lambda^j(k-1)),
\end{eqnarray} 
Note that by \eqref{w_update_ADMM} and \eqref{z_update_ADMM}, $W^i(k+1)=W^i(k)+2c\sum_{j\in N_i}\lambda^i(k)-\frac{\lambda^i(k)+\lambda^j(k)}{2}$. Then, 
We add the second Nash condition \eqref{Nash_equations} to \eqref{darakhar2}. Moreover, using Assumption~\ref{constraint}, there exists a matrix $B^i$ such that we multiply into \eqref{darakhar2} and then multiply the result by $(\lambda^i(k)-\lambda^*)$.
\begin{eqnarray}\label{OptimalLamndanahayi}
&&\hspace{-0.7cm}0=-(x_i^{i}(k)-x_i^*)^T{A_i^i}^T(\lambda^i(k)-\lambda^*)\nonumber\\
&&\hspace{-0.7cm}+\Big(W^i(k+1)-{W^i}^*\Big)^T{B^i}^T(\lambda^i(k)-\lambda^*)\nonumber\\
&&\hspace{-0.7cm}+c\sum_{j\in N_i}(\lambda^i(k)+\lambda^j(k)-\lambda^i(k-1)-\lambda^j(k-1))^T\nonumber\\
&&\hspace{-0.7cm}.{B^i}^T(\lambda^i(k)-\lambda^*).
\end{eqnarray}
 Adding \eqref{OptimalLamndanahayi} to \eqref{optimalx_inahayi}, it yields,
\begin{eqnarray}\label{optimalbeforx-i}
&&\hspace{-0.7cm}\Big(\nabla_iJ_i(x^i(k-1))-\nabla_iJ_i(x^*)\Big)^T(x_i^i(k-1)-x_i^*)\nonumber\\
&&\hspace{-0.7cm}+\Big(\nabla_iJ_i(x^i(k-1))-\nabla_iJ_i(x^*)\Big)^T(x_i^i(k)-x_i^i(k-1))\nonumber\\
&&\hspace{-0.7cm}+\Big(\partial_i\mathcal{I}_{\Omega_i}(x_i^i(k))-\partial_i\mathcal{I}_{\Omega_i}(x_i^*)\Big)^T(x_i^i(k)-x_i^*)\nonumber\\
&&\hspace{-0.7cm}+\delta_i(x_i^i(k)-x_i^i(k-1))^T(x_i^i(k)-x_i^*)\nonumber\\
&&\hspace{-0.7cm}+\Big(U_i^i(k)-{U_i^i}^*\Big)^T(x_i^i(k)-x_i^*)\nonumber\\
&&\hspace{-0.7cm}+\Big(W^i(k+1)-{W^i}^*)^T{B^i}^T(\lambda^i(k)-\lambda^*)\nonumber\\
&&\hspace{-0.7cm}+c\sum_{j\in N_i}(\lambda^i(k)+\lambda^j(k)-\lambda^i(k-1)-\lambda^j(k-1))^T.\nonumber\\
&&\hspace{-0.7cm}.{B^i}^T(\lambda^i(k)-\lambda^*)\\
&&\hspace{-0.7cm}+2c\!\sum_{j\in N_i}\!\Big(\!x_i^i(k)\!-\!\frac{x_i^i(k-1)\!+\!x_i^j(k-1)}{2}\Big)^T\!(x_i^i(k)\!-\!x_i^*)=0.\nonumber
\end{eqnarray}
The last equation that we need to add it to \eqref{optimalbeforx-i} is the one associated with $x_{-i}^i$, \eqref{x_-i_Update}, and multiplied by $(x_{-i}^i-x_{-i}^*)$.
\begin{eqnarray}\label{u_for_-i_*_mult}
&&\hspace{-1cm}U_{-i}^i(k)^T(x_{-i}^i(k)-x_{-i}^*)\nonumber\\
&&\hspace{-1cm}+2c\sum_{j\in N_i}\Big(x_{-i}^i(k)-\frac{x_{-i}^i(k-1)+x_{-i}^j(k-1)}{2}\Big)^T\nonumber\\
&&\hspace{-1cm}.(x_{-i}^i(k)-x_{-i}^*)=0.
\end{eqnarray}
Adding \eqref{u_for_-i_*_mult} to \eqref{optimalbeforx-i}, and using \eqref{u_update_ADMM} and \eqref{v_update_ADMM} for simplification, we obtain,
\begin{eqnarray}\label{optimalghablejam}
&&\hspace{-0.7cm}\Big(\nabla_iJ_i(x^i(k-1))-\nabla_iJ_i(x^*)\Big)^T(x_i^i(k-1)-x_i^*)\nonumber\\
&&\hspace{-0.7cm}+\Big(\nabla_iJ_i(x^i(k-1))-\nabla_iJ_i(x^*)\Big)^T(x_i^i(k)-x_i^i(k-1))\nonumber\\
&&\hspace{-0.7cm}+\Big(\partial_i\mathcal{I}_{\Omega_i}(x_i^i(k))-\partial_i\mathcal{I}_{\Omega_i}(x_i^*)\Big)^T(x_i^i(k)-x_i^*)\nonumber\\
&&\hspace{-0.7cm}+\delta_i(x_i^i(k)-x_i^i(k-1))^T(x_i^i(k)-x_i^*)\nonumber\\
&&\hspace{-0.7cm}+\!\sum_{j\in N_i}\!(u_i^{ij}(k+1)\!+\!v_i^{ji}(k+1)\!-\!{u_i^{ij}}^*\!-\!{v_i^{ji}}^*)^T\!(x_i^i(k)\!-\!x_i^*)\nonumber\\
&&\hspace{-0.7cm}+\sum_{j\in N_i}(u_{-i}^{ij}(k+1)+v_{-i}^{ji}(k+1))^T(x_{-i}^i(k)-x_{-i}^*)\nonumber\\
&&\hspace{-0.7cm}+\sum_{j\in N_i}(w^{ij}(k+1)+z^{ji}(k+1)-{w^{ij}}^*-{z^{ji}}^*)^T\nonumber\\
&&\hspace{-0.7cm}.{B^i}^T(\lambda^i(k)-\lambda^*)\nonumber\\
&&\hspace{-0.7cm}+c\sum_{j\in N_i}(\lambda^i(k)+\lambda^j(k)-\lambda^i(k-1)-\lambda^j(k-1))^T\nonumber\\
&&\hspace{-0.7cm}.{B^i}^T(\lambda^i(k)-\lambda^*).\nonumber\\
&&\hspace{-0.7cm}+c\sum_{j\in N_i}\Big(x^i(k)+x^j(k)-x^i(k-1)-x^j(k-1)\Big)^T\nonumber\\
&&\hspace{-0.7cm}.(x_i^i(k)-x_i^*)=0.
\end{eqnarray}
The second and third terms are bounded as follows:\vspace{-0cm}
\begin{eqnarray}\label{cauchy_Schewwrtz}
&&\hspace{-0.8cm}\Big(\nabla_iJ_i(x^i(k-1))\!-\!\nabla_iJ_i(x^*)\Big)^T\!(x_i^i(k)\!-\!x_i^i(k-1))\geq\\
&&\hspace{-0.8cm}\frac{-1}{2\rho}\!\|\!\nabla_iJ_i(x^i(k-1))\!-\!\nabla_iJ_i(x^*)\!\|^2\!-\!\frac{\rho}{2}\!\|\!x_i^i(k)\!-\!x_i^i(k-1)\!\|^2,\nonumber
\end{eqnarray}
for any $\rho>0$ $\forall i\in V$. By the convexity of $\mathcal{I}_{\Omega_i}$ (Assumption~\ref{assump}), we have for the third term,
\begin{equation}\label{convexity_of_Indicator}
(\partial\mathcal{I}_{\Omega_i}(x_i^i(k))-\partial\mathcal{I}_{\Omega_i}(x_i^*))^T(x_i^i(k)-x_i^*)\geq 0.
\end{equation}
Using \eqref{cauchy_Schewwrtz} and \eqref{convexity_of_Indicator} in \eqref{optimalghablejam} and summing over $i\in V$, we obtain,
\begin{eqnarray}\label{optim_equation_Nash_combo_multiplic_Revised_Sum}
&&\hspace{-0.7cm}\Big(F(\underline{x}(k-1))-F(\underline{x}^*)\Big)^T(x(k-1)-x^*)\nonumber\\
&&\hspace{-0.7cm}-\frac{1}{2\rho}\|F(\underline{x}(k-1))-F(\underline{x}^*)\|^2-\frac{1}{2}\|\underline{x}(k)-\underline{x}(k-1)\|_{M_1}^2\nonumber\\
&&\hspace{-0.7cm}+(\underline{x}(k)-\underline{x}(k-1))^T\text{diag}((\delta_ie_ie_i^T)_{i\in V})(\underline{x}(k)-\underline{x}^*)\nonumber\\
&&\hspace{-0.7cm}+\!\sum_{i\in V}\!\sum_{j\in N_i}\!(\!u_i^{ij}(k+1)\!+\!v_i^{ji}(k+1)\!-\!{u_i^{ij}}^*\!-\!{v_i^{ji}}^*\!)^T\!(\!x_i^i(k)\!-\!x_i^*\!)\nonumber\\
&&\hspace{-0.7cm}+\sum_{i\in V}\sum_{j\in N_i}(u_{-i}^{ij}(k+1)+v_{-i}^{ji}(k+1))^T(x_{-i}^i(k)-x_{-i}^*)\nonumber\\
&&\hspace{-0.7cm}+\frac{2}{c}(\underline{w}(k+1)-\underline{w}^*)^T\text{diag}(({B^i}^T)_{i\in V})(\underline{w}(k+1)-\underline{w}(k))\nonumber\\
&&\hspace{-0.7cm}+c(\underline{\lambda}(k)-\underline{\lambda}(k-1))^T((D+H)\otimes I_m)\nonumber\\
&&\hspace{-0.7cm}.\text{diag}(({B^i}^T)_{i\in V})(\underline{\lambda}(k)-\underline{\lambda}^*)+c(\underline{x}(k)-\underline{x}(k-1))^T\nonumber\\
&&\hspace{-0.7cm}.((D+H)\otimes I_N)(\underline{x}(k)-\underline{x}^*)\leq 0,
\end{eqnarray}
where $M_1:=\text{diag}((\rho e_ie_i^T)_{i\in V})$, $\underline{x}^*=[{{x^1}^*}^T,\ldots,{{x^N}^*}^T]^T$, $\underline{\lambda}=[{{\lambda^1}}^T,\ldots,{{\lambda^N}}^T]^T$ and $\underline{\lambda}^*=[{{\lambda^1}^*}^T,\ldots,{{\lambda^N}^*}^T]^T$. Moreover, $\underline{w}=(w^i)_{i\in V}\in \mathbb{R}^{m\sum_{i\in V}|N_i|}$ and $w^i=(w^{ij})_{j\in N_i}\in\mathbb{R}^{m|N_i|}$ and also $\underline{w}^*=({w^i}^*)_{i\in V}\in \mathbb{R}^{m\sum_{i\in V}|N_i|}$ and ${w^i}^*=({w_i^{ij}}^*)_{j\in N_i}\in\mathbb{R}^{m|N_i|}$.

We bound the first term using Assumption~\ref{Lip_assump},\vspace{-0.3cm}
\begin{eqnarray}\label{Cocoe_in_equat}
&&\hspace{-1cm}\Big(F(\underline{x}(k-1))-F(\underline{x}^*)\Big)^T(x(k-1)-x^*)\nonumber\\
&&\hspace{-1cm}\geq\sigma_F\|F(\underline{x}(k-1))-F(\underline{x}^*)\|^2.
\end{eqnarray}
We also simplify the fourth and the fifth lines in \eqref{optim_equation_Nash_combo_multiplic_Revised_Sum}. Since $G_C$ is an undirected graph, for any $\{a_{ij}\}$, $\sum_{i\in V}\sum_{j\in N_i}a_{ij}=\sum_{i\in V}\sum_{j\in N_i}a_{ji}$. Then,\vspace{-0.2cm}
\begin{eqnarray}\label{simpil_u_underline_bef}
&&\hspace{-0.8cm}\sum_{i\in V}\!\sum_{j\in N_i}\!(u_i^{ij}(k+1)\!+\!v_i^{ji}(k+1)\!-\!{u_i^{ij}}^*\!-\!{v_i^{ji}}^*)^T\!(x_i^i(k)\!-\!x_i^*)\nonumber\\
&&\hspace{-0.8cm}+\sum_{i\in V}\sum_{j\in N_i}(u_{-i}^{ij}(k+1)+v_{-i}^{ji}(k+1))^T(x_{-i}^i(k)-x_{-i}^*)\nonumber\\
&&\hspace{-0.8cm}=\sum_{i\in V}\sum_{j\in N_i}(u_i^{ij}(k+1)-{u_i^{ij}}^*)^T(x_i^i(k)-x_i^*)\nonumber\\
&&\hspace{-0.8cm}+\sum_{i\in V}\sum_{j\in N_i}(v_i^{ij}(k+1)-{v_i^{ij}}^*)^T(x_i^j(k)-x_i^*)\nonumber\\
&&\hspace{-0.8cm}+\sum_{i\in V}\sum_{j\in N_i}u_{-i}^{ij}(k+1)^T(x_{-i}^i(k)-x_{-i}^*)\nonumber\\
&&\hspace{-0.8cm}+\sum_{i\in V}\sum_{j\in N_i}v_{-i}^{ij}(k+1)^T(x_{-i}^j(k)-x_{-i}^*).
\end{eqnarray}
Note that by \eqref{u_update_ADMM} and \eqref{v_update_ADMM} as well as the initial conditions for Lagrange multipliers $u^{ij}(0)=v^{ij}(0)=\textbf{0}_N$ $\forall i\in V,\,j\in N_i$, we obtain, \begin{equation}\label{u+v=0}u^{ij}(k)+v^{ij}(k)=\textbf{0}_N\quad \forall i\in V,\,j\in N_i,\,k>0.
\end{equation} 
Substituting \eqref{u+v=0} into \eqref{simpil_u_underline_bef} and using \eqref{Nash_equations}, we obtain,\vspace{-0.2cm}
\begin{eqnarray}\label{simpil_u_underline}
&&\hspace{-0.8cm}\sum_{i\in V}\sum_{j\in N_i}(u_i^{ij}(k+1)-{u_i^{ij}}^*)^T(x_i^i(k)-x_i^j(k))\nonumber\\
&&\hspace{-0.8cm}+\sum_{i\in V}\sum_{j\in N_i}u_{-i}^{ij}(k+1)^T(x_{-i}^i(k)-x_{-i}^j(k))\nonumber\\
&&\hspace{-0.8cm}=\frac{2}{c}\sum_{i\in V}\sum_{j\in N_i}(u^{ij}(k+1)-{u_i^{ij}}^*e_i)^T(u^{ij}(k+1)-{u^{ij}}(k))\nonumber\\
&&\hspace{-0.8cm}:=\frac{2}{c}(\underline{u}(k+1)-\underline{u}^*)^T(\underline{u}(k+1)-\underline{u}(k)),
\end{eqnarray}
where $\underline{u}=(u^i)_{i\in V}\in \mathbb{R}^{N\sum_{i\in V}|N_i|}$ and $u^i=(u^{ij})_{j\in N_i}\in\mathbb{R}^{N|N_i|}$ and also $\underline{u}^*=({u^i}^*)_{i\in V}\in \mathbb{R}^{N\sum_{i\in V}|N_i|}$ and ${u^i}^*=({u_i^{ij}}^*)_{j\in N_i}\otimes e_i\in\mathbb{R}^{N|N_i|}$. Using \eqref{Cocoe_in_equat} and \eqref{simpil_u_underline}, for $\rho=\frac{1}{2\sigma_F}$ \eqref{optim_equation_Nash_combo_multiplic_Revised_Sum} becomes,\vspace{-0.3cm}
\begin{eqnarray}\label{optim_equation_Nash_combo_multiplic_Revised_Sum_simpil}
&&\hspace{-0.8cm}-\frac{1}{2}\|\underline{x}(k)-\underline{x}(k-1)\|_{M_1}^2\nonumber\\
&&\hspace{-0.8cm}+(\underline{x}(k)-\underline{x}(k-1))^TM_2(\underline{x}(k)-\underline{x}^*)\nonumber\\
&&\hspace{-0.8cm}+(\underline{\lambda}(k)-\underline{\lambda}(k-1))^TQ(\underline{\lambda}(k)-\underline{\lambda}^*)\nonumber\\
&&\hspace{-0.8cm}+\frac{2}{c}(\underline{u}(k+1)-\underline{u}^*)^T(\underline{u}(k+1)-\underline{u}(k))\nonumber\\
&&\hspace{-0.8cm}+\frac{2}{c}(\underline{w}(k+1)-\underline{w}^*)^TB(\underline{w}(k+1)-\underline{w}(k))\leq 0,
\end{eqnarray}
where $M_2:=\text{diag}\Big((\delta_i e_ie_i^T)_{i\in V}\Big)+c((D+H)\otimes I_N)$, $B:=\text{diag}(({B^i}^T)_{i\in V})$ and $Q:=c((D+H)\otimes I_N)B$. Note that,
\begin{equation}\label{D+A_L}
c((D+A)\otimes I_N)=c((D^{\frac{1}{2}}(2I-L_{N})D^{\frac{1}{2}})\otimes I_N),
\end{equation} 
where $L:=D-H$, $D^{\frac{1}{2}}$, $D^{-\frac{1}{2}}$ and $L_N:=D^{-\frac{1}{2}}LD^{-\frac{1}{2}}$ are the Laplacian of $G_C$, the square root and reciprocal square root of $D$ and the normalized Laplacian of $G_C$, respectively. Since $D\succ 0$, $D^{-\frac{1}{2}}$ exist, it is shown in 
that $\lambda_{\max}(L_N)\leq2$. Then \eqref{D+A_L} yields that $c((D+H)\otimes I_N)\succeq0$. This concludes $M_2\succeq0$. Moreover, by Assumption~\ref{constraint}, $B\succeq 0$. We use the following inequality in \eqref{optim_equation_Nash_combo_multiplic_Revised_Sum_simpil} for every $\{a(k)\}$ and $M\succeq 0$:
\begin{eqnarray}
&&\hspace{-1cm}(a(k)-a(k-1))^TM(a(k)-a^*)=\frac{1}{2}\|a(k)-a^*\|_M^2\nonumber\\
&&\hspace{-1cm}+\frac{1}{2}\|a(k)-a(k-1)\|_M^2-\frac{1}{2}\|a(k-1)-a^*\|_M^2.
\end{eqnarray}
Then \eqref{optim_equation_Nash_combo_multiplic_Revised_Sum_simpil} becomes,
\begin{eqnarray}\label{optim_equation_Nash_combo_multiplic_Revised_Sum_simpil_square}
&&\hspace{-0.8cm}\Big\{\frac{1}{2}\|\underline{x}(k)-\underline{x}^*\|^2_{M_2}+\frac{1}{2}\|\underline{\lambda}(k)-\underline{\lambda}^*\|^2_{Q}\nonumber\\
&&\hspace{-0.8cm}+\frac{1}{c}\|\underline{u}(k+1)-\underline{u}^*\|^2+\frac{1}{c}\|\underline{w}(k+1)-\underline{w}^*\|_B^2\Big\}\leq\nonumber\\
&&\hspace{-0.8cm}\Big\{\frac{1}{2}\|\underline{x}(k-1)-\underline{x}^*\|^2_{M_2}+\frac{1}{2}\|\underline{\lambda}(k-1)-\underline{\lambda}^*\|^2_{Q}\nonumber\\
&&\hspace{-0.8cm}+\frac{1}{c}\|\underline{u}(k)-\underline{u}^*\|^2+\frac{1}{c}\|\underline{w}(k)-\underline{w}^*\|_B^2\Big\}\nonumber\\
&&\hspace{-0.8cm}-\frac{1}{2}\|\underline{x}(k)-\underline{x}(k-1)\|_{M_2-M_1}^2-\frac{1}{2}\|\underline{\lambda}(k)-\underline{\lambda}(k-1)\|^2_Q\nonumber\\
&&\hspace{-0.8cm}-\frac{1}{c}\|\underline{u}(k+1)-\underline{u}(k)\|^2-\frac{1}{c}\|\underline{w}(k+1)-\underline{w}(k)\|_B^2.
\end{eqnarray}
By the condition \eqref{condition}, $M_2-M_1\succ0$. Then \eqref{optim_equation_Nash_combo_multiplic_Revised_Sum_simpil_square} implies the following two results:
\begin{enumerate}
	\item $\frac{1}{2}\|\underline{x}(k)-\underline{x}^*\|^2_{M_2}+\frac{1}{2}\|\underline{\lambda}(k)-\underline{\lambda}^*\|^2_{Q}+\frac{1}{c}\|\underline{u}(k+1)-\underline{u}^*\|^2+\frac{1}{c}\|\underline{w}(k+1)-\underline{w}^*\|_B^2\rightarrow\theta$, for some $\theta\geq0$,
	\vspace{0.3cm}
	\item $\begin{cases}\underline{x}(k)-\underline{x}(k-1)\rightarrow\textbf{0}_{N^2}\\
	\underline{\lambda}(k)-\underline{\lambda}(k-1)\rightarrow\textbf{0}_{mN}\\
	\underline{u}(k+1)-\underline{u}(k)\rightarrow\textbf{0}_{N\sum_{i\in V}|N_i|}\\
	\underline{w}(k+1)-\underline{w}(k)\rightarrow\textbf{0}_{m\sum_{i\in V}|N_i|}\end{cases}$.
\end{enumerate}
Result 1 implies that the sequences $\{x^i(k)\},\,\{\lambda^i(k)\},\,\{u^{ij}(k)\}$ and $\{w^{ij}(k)\}$ (similarly $\{v^{ij}(k)\}$ and $\{z^{ij}(k)\}$) are bounded and have limit points denoted by $\bar{x}^i,\,\bar{\lambda}^i,\,\bar{u}^{ij}$ and $\bar{w}^{ij}$ ($\bar{v}^{ij}$ and $\bar{z}^{ij}$), respectively. Then, we obtain,
\begin{eqnarray}\label{theta}
&&\hspace{-0.8cm}\theta=\frac{1}{2}\|\underline{\bar{x}}(k)-\underline{x}^*\|^2_{M_2}+\frac{1}{2}\|\underline{\bar{\lambda}}(k)-\underline{\lambda}^*\|^2_{Q}\nonumber\\
&&\hspace{-0.8cm}+\frac{1}{c}\|\underline{\bar{u}}(k+1)-\underline{u}^*\|^2+\frac{1}{c}\|\underline{\bar{w}}(k+1)-\underline{\bar{w}}^*\|_B^2
\end{eqnarray}
Result 2 along with \eqref{u_update_ADMM} and \eqref{w_update_ADMM} yield $\forall i\in V,\,j\in N_i$, 
\begin{equation}\label{xi=xjlambdai=lambdaj}
\bar{x}^i=\bar{x}^j:=\bar{x},\,\bar{\lambda}^i=\bar{\lambda}^j:=\bar{\lambda}
\end{equation}
Moreover, by \eqref{u+v=0} we arrive at,
\begin{equation}\label{u+v=0tilda}\bar{u}^{ij}+\bar{v}^{ij}=\textbf{0}_N\quad \forall i\in V,\,j\in N_i.\end{equation}
Similarly, $\bar{w}^{ij}+\bar{z}^{ij}=\textbf{0}_m\quad \forall i\in V,\,j\in N_i$. Result 2 also implies that by \eqref{xi=xjlambdai=lambdaj}, \eqref{darakhar1} and \eqref{darakhar2}:
\begin{eqnarray}
&&\nabla_iJ_i(\bar{x})+\partial\mathcal{I}_{\Omega_i}(\bar{x}_i)+\sum_{j\in N_i}(\bar{u}_i^{ij}+\tilde{v}_i^{ji})=0,\label{optim_equation_tilda}\\
&&A^i\bar{x}-b^i-\sum_{j\in N_i}{\bar{w}^{ij}}+{\bar{z}^{ji}}=\textbf{0}_m.\label{shabihnash}
\end{eqnarray}
Comparing \eqref{xi=xjlambdai=lambdaj}-\eqref{shabihnash} with \eqref{Nash_equations}, it follows $\forall i\in V,\,j\in N_i$,
\begin{eqnarray}
&&\hspace{-0.8cm}\bar{x}^i=x^*\,(\underline{\bar{x}}=\underline{x^*}),\,\bar{\lambda}^i=x^*\,(\underline{\bar{\lambda}}=\underline{\lambda^*}),\label{xxlambdatilde=lambdastar}\\
&&\hspace{-0.8cm}\bar{u}^{ij}={u^{ij}}^*\,(\underline{\bar{u}}=\underline{u^*}),\,\bar{w}^{ij}={w^{ij}}^*\,(\underline{\bar{w}}=\underline{w^*}).\label{wtilde=wstar}
\end{eqnarray} 
This completes the proof.
$\hfill\blacksquare$
\section{Simulation Results}
In this section, we simulate our algorithm for a wireless ad-hoc network (WANET). Consider a WANET with 16 nodes and 16 multi-hop communication links as in Fig.~1~(a). There are 15 users who aim to transfer data from a source to a destination. Solid line represents a link $L_j,\,j\in\{1,\ldots,16\}$ and dashed line displays a path $R_i,\,i\in\{1,\ldots,15\}$ that is assigned to user $i$ to transfer data. Each link $L_j$ has a positive capacity $C_j>0$ that restricts the users' data flow .
\begin{figure}
	\vspace{-1.5cm}
	\hspace{-2.23cm}
	\centering
	\includegraphics [scale=0.4]{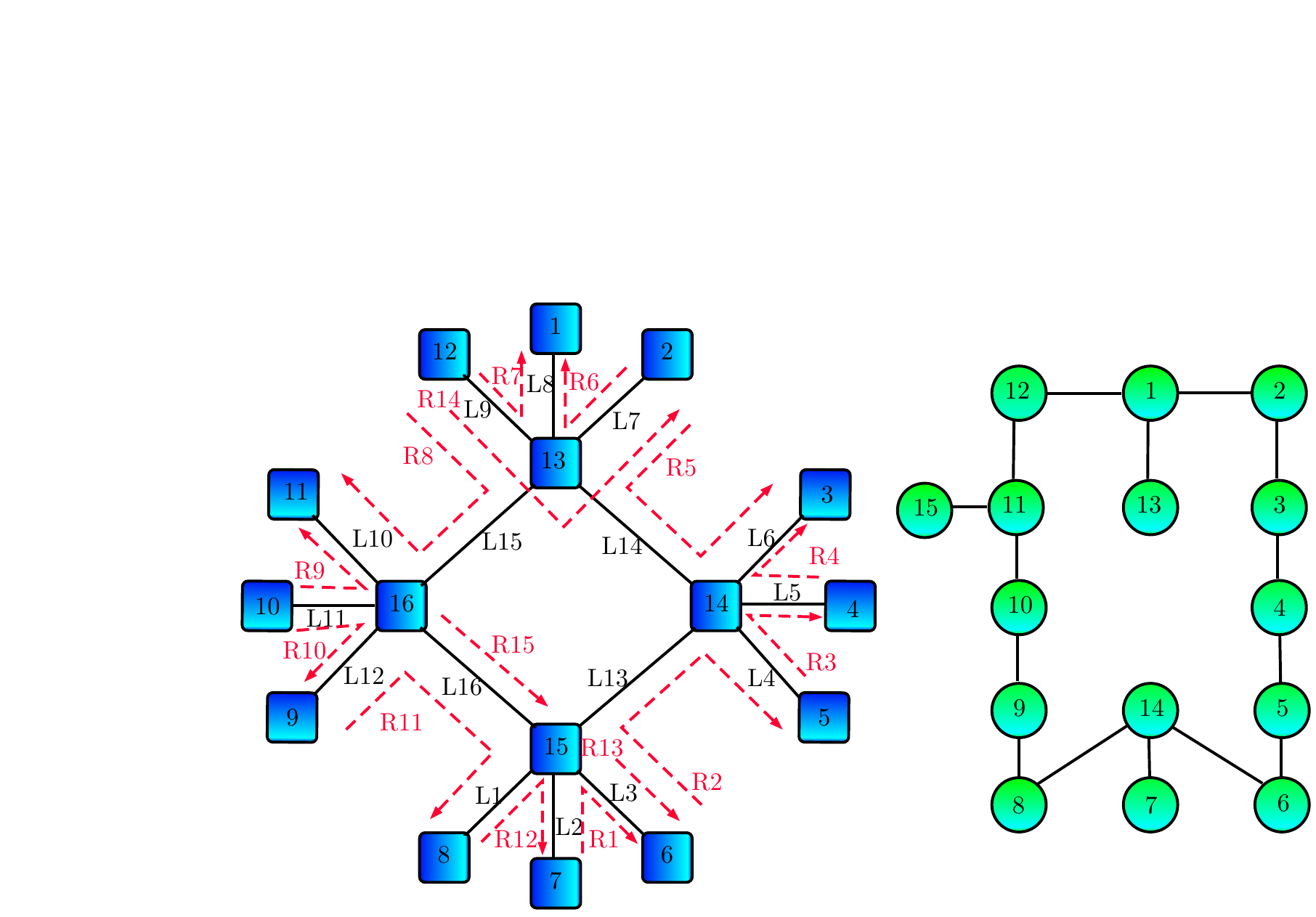}
	\vspace{-0.1cm}
	\caption{(a)  Wireless Ad-Hoc Network (left). (b) Communication graph $G_C$ (right).}
\end{figure}
The data flow of user $i$ is denoted by $x_i$ such that $0\leq x_i\leq 10$. 
For each user $i$, a cost function $J_i$ is defined as in \cite{ssalehisadaghiani2016distributed}:
\begin{equation*}\label{Cost_fcn_gen}
J_i(x_i,x_{-i}):=\sum_{j:L_j\in R_i}\frac{\kappa}{C_j-\sum_{w:L_j\in R_w}x_w}-\chi_i \log(x_i+1),\end{equation*}
where $\kappa>0$ and $\chi_i>0$ are network-wide known and user-specific parameters, respectively. 

The problem is to find a GNE of the game which is played over a communication graph $G_C$ (depicted in Fig.~1~(b)). We assume a common constraint $x_1+x_3+x_5=14$ between the players. It is straightforward to check the Assumptions 2,3 and 4 on $G_C$ and the cost functions. The users' flow rate and the dual Lagrange multipliers (they converge to 0) as well as the normalized error are shown in Fig.~2 and Fig.~3 for $\chi_i=15$ $\forall i\in\{1,\ldots,15\}$ and $C_j=15$ $\forall j\in\{1,\ldots,16\}$.
\begin{figure}
	\vspace{-2.8cm}
	\hspace{0.8cm}
	\includegraphics [scale=0.3]{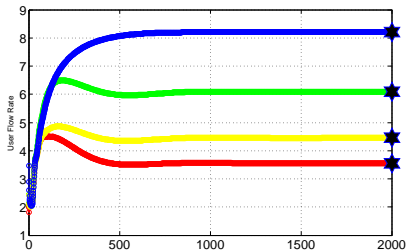}
	\label{fig:minipage1}
	\vspace{-2.6cm}
	\caption{Flow rates of users 1, 3, 5, 15. GNE points are represented by black stars.}
	\vspace{-0.6cm}
\end{figure}
\begin{figure}
	\vspace{-2.8cm}
	\hspace{0.8cm}
	\includegraphics [scale=0.3]{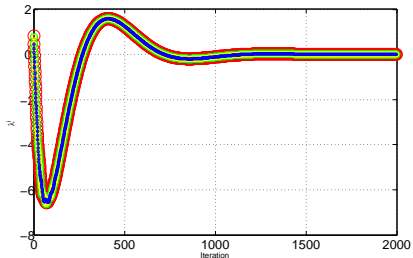}
	\label{fig:minipage1}
	\vspace{-2.6cm}
	\caption{Dual Lagrange multipliers for users 1, 3, 5, 15.}
	\vspace{-0.6cm}
\end{figure}
\begin{figure}
	\vspace{-2.5cm}
	\hspace{0.8cm}
	\includegraphics [scale=0.3]{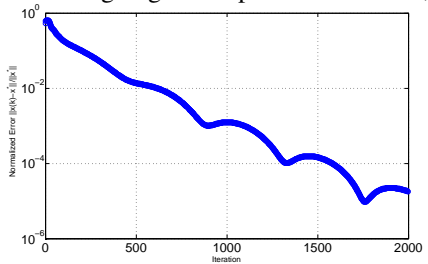}
	\label{fig:minipage1}
	\vspace{-2.5cm}
	\caption{Normalized error of the users' data flow.}
		\vspace{-0.5cm}
\end{figure}
\vspace{-0.2cm}
\section{Conclusions}
A distributed GNE seeking algorithm is designed within the framework of inexact-ADMM. Each player is only aware of his own cost function, problem data and action set of all players. Each player exchanges the estimates of other players' actions along with the local copy of his Lagrange multiplier with his communication neighbors. An inexact-ADMM algorithm is then derived to find a GNE of the game. Convergence of the algorithm is then provided under mild conditions on cost functions.
\vspace{-0.3cm}
\bibliographystyle{IEEEtran}
\bibliography{IEEEabrv,ref}

 \newcommand{\noop}[1]{}
\begin{thebibliography}{10}
\providecommand{\url}[1]{#1}
\csname url@samestyle\endcsname
\providecommand{\newblock}{\relax}
\providecommand{\bibinfo}[2]{#2}
\providecommand{\BIBentrySTDinterwordspacing}{\spaceskip=0pt\relax}
\providecommand{\BIBentryALTinterwordstretchfactor}{4}
\providecommand{\BIBentryALTinterwordspacing}{\spaceskip=\fontdimen2\font plus
\BIBentryALTinterwordstretchfactor\fontdimen3\font minus
  \fontdimen4\font\relax}
\providecommand{\BIBforeignlanguage}[2]{{%
\expandafter\ifx\csname l@#1\endcsname\relax
\typeout{** WARNING: IEEEtran.bst: No hyphenation pattern has been}%
\typeout{** loaded for the language `#1'. Using the pattern for}%
\typeout{** the default language instead.}%
\else
\language=\csname l@#1\endcsname
\fi
#2}}
\providecommand{\BIBdecl}{\relax}
\BIBdecl

\bibitem{zhu2016distributed}
M.~Zhu and E.~Frazzoli, ``Distributed robust adaptive equilibrium computation
  for generalized convex games,'' \emph{Automatica}, vol.~63, pp. 82--91, 2016.

\bibitem{jayash10}
L.~Pavel, ``An extension of duality to a game-theoretic framework,''
  \emph{Automatica}, vol.~43, no.~2, pp. 226--237, 2007.

\bibitem{han2012game}
Z.~Han, \emph{Game theory in wireless and communication networks: theory,
  models, and applications}.\hskip 1em plus 0.5em minus 0.4em\relax Cambridge
  University Press, 2012.

\bibitem{jayash8}
T.~Alpcan and T.~Ba{\c{s}}ar, ``Distributed algorithms for {N}ash equilibria of
  flow control games,'' in \emph{Advances in Dynamic Games}.\hskip 1em plus
  0.5em minus 0.4em\relax Springer, 2005, pp. 473--498.

\bibitem{facchinei2010generalized}
F.~Facchinei and C.~Kanzow, ``Generalized {Nash} equilibrium problems,''
  \emph{Annals of Operations Research}, vol. 175, no.~1, pp. 177--211, 2010.

\bibitem{fischer2014generalized}
A.~Fischer, M.~Herrich, and K.~Sch{\"o}nefeld, ``Generalized {Nash} equilibrium
  problems-recent advances and challenges,'' \emph{Pesquisa Operacional},
  vol.~34, no.~3, pp. 521--558, 2014.

\bibitem{ssalehisadaghiani2016distributed}
F.~Salehisadaghiani and L.~Pavel, ``Distributed {Nash} equilibrium seeking by
  gossip in games on graphs,'' in \emph{Decision and Control (CDC), 2016 IEEE
  55th Conference on}.\hskip 1em plus 0.5em minus 0.4em\relax IEEE, 2016, pp.
  6111--6116.

\bibitem{chang2015multi}
T.-H. Chang, M.~Hong, and X.~Wang, ``Multi-agent distributed optimization via
  inexact consensus admm,'' \emph{IEEE Transactions on Signal Processing},
  vol.~63, no.~2, pp. 482--497, 2015.

\bibitem{johansson2008distributed}
B.~Johansson, ``On distributed optimization in networked systems,'' \emph{Ph.D.
  Dissertation}, 2008.

\bibitem{parise2015network}
F.~Parise, B.~Gentile, S.~Grammatico, and J.~Lygeros, ``Network aggregative
  games: Distributed convergence to {Nash} equilibria,'' in \emph{2015 54th
  IEEE Conference on Decision and Control (CDC)}.\hskip 1em plus 0.5em minus
  0.4em\relax IEEE, 2015, pp. 2295--2300.

\bibitem{salehisadaghiani2016distributed}
F.~Salehisadaghiani and L.~Pavel, ``Distributed {Nash} equilibrium seeking: A
  gossip-based algorithm,'' \emph{Automatica}, vol.~72, pp. 209--216, 2016.

\bibitem{yi2017distributed}
P.~Yi and L.~Pavel, ``A distributed primal-dual algorithm for computation of
  generalized nash equilibria with shared affine coupling constraints via
  operator splitting methods,'' \emph{arXiv preprint arXiv:1703.05388}, 2017.

\bibitem{yin2011nash}
H.~Yin, U.~V. Shanbhag, and P.~G. Mehta, ``{Nash} equilibrium problems with
  scaled congestion costs and shared constraints,'' \emph{IEEE Transactions on
  Automatic Control}, vol.~56, no.~7, pp. 1702--1708, 2011.

\bibitem{schiro2013solution}
D.~A. Schiro, J.-S. Pang, and U.~V. Shanbhag, ``On the solution of affine
  generalized {Nash} equilibrium problems with shared constraints by lemke’s
  method,'' \emph{Mathematical Programming}, pp. 1--46, 2013.

\bibitem{paccagnan2016distributed}
D.~Paccagnan, B.~Gentile, F.~Parise, M.~Kamgarpour, and J.~Lygeros,
  ``Distributed computation of generalized {Nash} equilibria in quadratic
  aggregative games with affine coupling constraints,'' in \emph{Decision and
  Control (CDC), 2016 IEEE 55th Conference on}.\hskip 1em plus 0.5em minus
  0.4em\relax IEEE, 2016, pp. 6123--6128.

\bibitem{grammatico2016aggregative}
S.~Grammatico, ``Aggregative control of competitive agents with coupled
  quadratic costs and shared constraints,'' in \emph{Decision and Control
  (CDC), 2016 IEEE 55th Conference on}.\hskip 1em plus 0.5em minus 0.4em\relax
  IEEE, 2016, pp. 3597--3602.

\bibitem{bertsekas1999parallel}
D.~P. Bertsekas and J.~N. Tsitsiklis, \emph{Parallel and Distributed
  Computation: Numerical Methods}.\hskip 1em plus 0.5em minus 0.4em\relax
  Athena Scientific, Belmont, Massachusetts, 1997.

\bibitem{boyd2011distributed}
S.~Boyd, N.~Parikh, E.~Chu, B.~Peleato, and J.~Eckstein, ``Distributed
  optimization and statistical learning via the alternating direction method of
  multipliers,'' \emph{Foundations and Trends{\textregistered} in Machine
  Learning}, vol.~3, no.~1, pp. 1--122, 2011.

\bibitem{he20121}
B.~He and X.~Yuan, ``On the o(1/n) convergence rate of the douglas-rachford
  alternating direction method,'' \emph{SIAM Journal on Numerical Analysis},
  vol.~50, no.~2, pp. 700--709, 2012.

\bibitem{wei2012distributed}
E.~Wei and A.~Ozdaglar, ``Distributed alternating direction method of
  multipliers,'' in \emph{2012 IEEE 51st IEEE Conference on Decision and
  Control (CDC)}.\hskip 1em plus 0.5em minus 0.4em\relax IEEE, 2012, pp.
  5445--5450.

\bibitem{salehisadaghiani2016distributedifac}
F.~Salehisadaghiani and L.~Pavel, ``Distributed {Nash} equilibrium seeking via
  the alternating direction method of multipliers,'' \emph{accepted in IFAC
  World Congress}, 2017.

\end{thebibliography}
\end{document}